\documentclass{article}
\usepackage{spconf,amsmath,graphicx,hyperref,booktabs}
\usepackage{marginnote}
\usepackage{setspace} 
\usepackage{comment}

\title{Speech-Based Prioritization for Schizophrenia Intervention}
\name{Gowtham Premananth$^{1}$, Philip Resnik$^{2}$, Sonia Bansal$^{3}$, Deanna L.Kelly$^{3}$, Carol Espy-Wilson$^{1}$\thanks{This work was supported by the National Science Foundation grant numbered 2124270.}}
\address{$^{1}$ Institute for System Research, Department of Electrical and Computer Engineering,\\  
University of Maryland, College Park, USA\\
$^{2}$Institute for Advanced Computer Studies, University of Maryland, College Park, USA\\
$^{3}$School of Medicine, University of Maryland, USA}
\begin{document}
%
\maketitle
\begin{abstract}
Millions of people suffer from mental health conditions, yet many remain undiagnosed or receive delayed care due to limited clinical resources and labor-intensive assessment methods. While most machine-assisted approaches focus on diagnostic classification, estimating symptom severity is essential for prioritizing care, particularly in resource-constrained settings. Speech-based AI provides a scalable alternative by enabling automated, continuous, and remote monitoring, reducing reliance on subjective self-reports and time-consuming evaluations. In this paper, we introduce a speech-based model for pairwise comparison of schizophrenia symptom severity, leveraging articulatory and acoustic features. These comparisons are used to generate severity rankings via the Bradley-Terry model. Our approach outperforms previous regression-based models on ranking-based metrics, offering a more effective solution for clinical triage and prioritization.
\end{abstract}
\vspace{-0.3cm}
\section{Introduction}
\vspace{-0.3cm}
Mental health conditions affect millions globally \cite{institute2021global}, affecting not only individuals but also the ones around them. While advancements in medical science have led to effective intervention strategies such as medication, therapy, and lifestyle modifications, timely and accurate assessment and diagnosis remain a critical challenge. Many individuals go undiagnosed or receive delayed care due to limited access to healthcare resources: in the U.S., more than a third of the population live in federally designated Mental Health Professional Shortage Areas \cite{HRSA2024}. This underscores the need for intelligent systems that can identify individuals requiring clinical intervention and assist healthcare professionals in prioritizing care.

The rise of artificial intelligence (AI) and deep learning has led to the development of AI-driven early diagnosis systems \cite{early_cancer_diagnosis} and remote healthcare monitoring solutions \cite{remotemonitoring}. However, these models often require large, high-quality datasets, which are scarce in the medical field due to privacy constraints. This limitation is particularly challenging when the task extends beyond classification to severity estimation, which is essential for intervention prioritization.

Among various AI-driven approaches, speech-based analysis offers a non-invasive, scalable, and cost-effective solution for mental health assessment. Speech carries valuable information related to cognitive, emotional, and neurological states, supporting AI based approaches as a promising tool for measuring symptoms severity in conditions like depression \cite{seneviratne21b_dep}, schizophrenia \cite{premananth25_interspeech},  and post traumatic stress disorder \cite{kathan24_ptsd} . Unlike traditional diagnostic methods, speech-based AI models enable automated, remote-friendly, and continuous monitoring, making early intervention more accessible, particularly in underserved regions. 

Most AI-based mental health studies focus on classification, determining whether an individual has a disorder or specific symptom \cite{seneviratne2022multimodal,embc,premananth24_schizo}. However, in clinical practice, diagnosis alone is often insufficient, as patients with the same condition can vary widely in symptom severity and care needs. A more critical need is to rank individuals by psychiatric symptom severity to help clinicians prioritize those requiring urgent intervention. Estimating severity enables a more nuanced understanding of patient needs, allowing for better resource allocation and more timely treatment. Integrating severity assessment into AI systems moves beyond binary classification toward a dynamic, patient-centered model that improves overall care management.

In this study, we introduce a speech-based approach for schizophrenia intervention prioritization using symptom severity. Using articulatory and acoustic speech features, we develop a pairwise comparison prediction model, whose outputs are fed into the Bradley-Terry model \cite{bradley1952rank} to generate a severity ranking. By framing severity estimation as a ranking problem rather than a binary classification task, this method provides a more nuanced and clinically relevant assessment, aiding healthcare providers in prioritizing urgent cases and improving patient outcomes.

The key contributions of this paper are:
\vspace{-0.2cm}
\begin{enumerate}
    \item A speech-based pairwise comparison prediction model leveraging Vocal Tract Variable(TV)-based articulatory features and pre-trained self-supervised acoustic features to determine which of two sessions exhibits higher severity.
    \vspace{-0.3cm}
    \item A Bradley-Terry algorithm-based ranking framework that processes pairwise comparison predictions to rank sessions by symptom severity, aiding clinicians in prioritizing patients especially when resources are limited.
\end{enumerate}

\begin{figure*}[th!]
  \centering
  \includegraphics[width=0.9\linewidth]{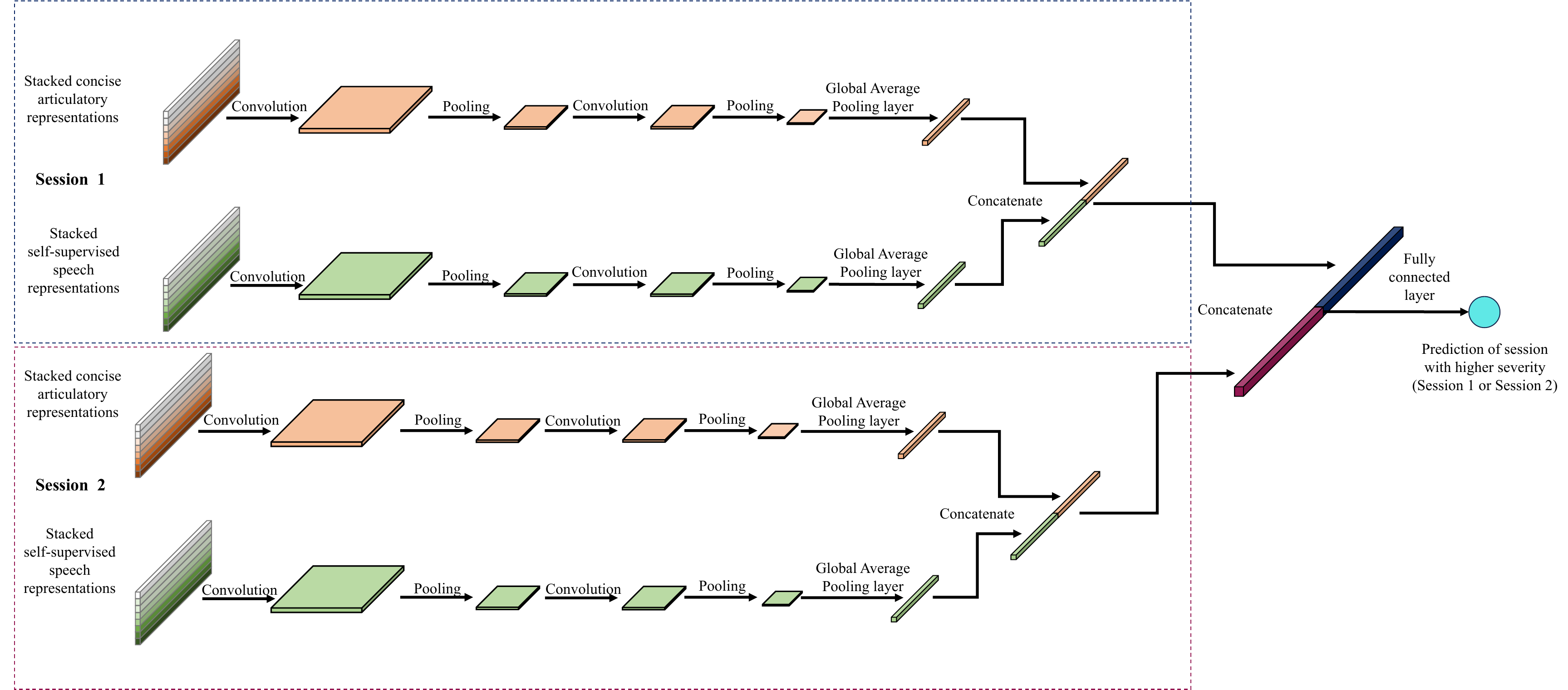}
  \caption{\textbf{Speech-based Pairwise comparison prediction model with acoustic and articulation fusion approach }\footnotesize{(The model architecture consists of only two branches: acoustic and articulatory. Inputs from both sessions are processed through these same branches. The figure shows four branches simply to illustrate that inputs from two sessions pass through the acoustic and articulatory branches, after which their outputs are fused to generate the pairwise comparison prediction.)}}
  \label{fig:pairwise}
\end{figure*}

\vspace{-0.6cm}
\section{Method}
\vspace{-0.3cm}
This section outlines the dataset, feature extraction, and the models and algorithms used in this study.
\vspace{-0.3cm}
\subsection{Dataset}
\vspace{-0.2cm}
The dataset used in this study was collected through a collaborative effort between the University of Maryland School of Medicine and the University of Maryland, College Park, as part of research on assessing mental health disorders such as major depressive disorder and schizophrenia \cite{Kelly2020-cj}. It consists of audio and video recordings of in-person, unstructured clinical interviews conducted over time, in which interviewers engaged subjects in open-ended conversations guided by their spontaneous responses rather than a predefined set of questions. In addition to these recordings, the dataset includes clinical evaluation scores obtained prior to each session.

For this work, we focused on a subset of the dataset comprising 140 sessions from 39 unique subjects, including both healthy controls and subjects with schizophrenia. Only the audio recordings were used, and symptom severity was measured using evaluation scores based on the 18-item Brief Psychiatric Rating Scale (BPRS) \cite{overall1962brief}. The BPRS, widely regarded as a gold standard in psychiatry, provides a validated measure of symptom severity in schizophrenia. In the sessions used for this study, BPRS scores ranged from 19 to 62.

\vspace{-0.3cm}
\subsection{Feature extraction}
\vspace{-0.2cm}
The audio recordings in the dataset consist of speech from both the interviewer and the subject. Therefore initially the speech from the session was diarized and then segmented into 40-second segments. The 40-second segments were used for articulatory and acoustic feature extraction. For articulatory features, a total of six Vocal Tract Variables (TVs) that represent the temporal and spatial movements of the articulators in the vocal tract (lips, tongue tip, and tongue body) were extracted using an acoustic-to-articulatory speech inversion system\cite{speechinversion}. In addition to the TVs, articulatory source features aperiodicty and periodicity were extracted using an Aperioidicity Periodicity Pitch detector \cite{appdetector}. From the combination of TVs and source features, a Full Vocal Tract Coordination (FVTC) structure that portrays phasing between articulatory gestures was calculated using channel-delay correlations \cite{FVTC}. As previous work on speech-based schizophrenia severity estimation \cite{gowtham_spade} has shown that concise articulatory representations extracted from FVTCs, using a Vector Quantized-Variational Auto Encoder (VQ-VAE) \cite{van2017neural} based representation learning model has outperformed the use of stacked FVTCs, we also used concise articulatory representations as articulatory features for all our models. We extracted these concise representations from the same model introduced in \cite{gowtham_spade}. Acoustic features were extracted using a pre-trained  Wav2Vec2.0 \cite{wav2vec} self-supervised model.

\vspace{-0.3cm}
\subsection{Model for speech-based pairwise comparison}
\vspace{-0.2cm}
The pairwise comparison prediction model leverages both articulatory and acoustic representations to assess symptom severity. For each 40-second segment within a session, articulatory features of length 1024 are extracted and stacked to create session-level articulatory representations, while 768-dimensional Wav2Vec2.0 embeddings are similarly stacked to form session-level acoustic speech representations. The model takes inputs from two sessions and predicts a binary outcome indicating which session corresponds to a higher severity level, with ground-truth labels derived from comparisons of the sessions’ BPRS scores \footnote{We note that although we used the total BPRS score in this study, it would be entirely straightforward to use scores associated with sub-scales of the BPRS.}. Session pairs with identical BPRS scores are excluded from training; however, their individual sessions still contribute through comparisons with other sessions. This design ensures that all sessions can still be evaluated and strengthens the model’s robustness by leveraging population-level comparisons for more reliable severity estimation.

The model architecture as shown in Fig.\ref{fig:pairwise} uses a convolution neural network architecture. For the inputs from each session both the acoustic features and articulatory features are sent through their respective branches that comprise of 2 sets of convolution layers, batch normalization layers and pooling layers. These are then then sent through a global average pooling layer to convert the 2D matrices to vectors before fusion. Then the acoustic and articulatory branches are fused together using a simple concatenation. After obtaining the fused representations of both the sessions, they are sent through a set of fully connected layers before making the final binary prediction of which session has the higher BPRS score, i.e. the greater symptom severity.

\vspace{-0.3cm}
\subsection{Ranking based on pairwise comparisons}
\vspace{-0.2cm}
The Bradley-Terry model \cite{bradley1952rank} is a statistical framework designed to analyze the outcomes of pairwise comparisons. It operates under the assumption that each pairwise comparison is independent of the others. Each item in a comparison population (here, a session) has a latent ``strength'' score (here, the BPRS severity for the individual in the session). Viewed generatively, the model predicts the probability that one item will rank higher than the other item in the ranking of items. Expressed in terms of sessions $S_1, S_2$:
\begin{equation}
    P(Score_{S_1}>Score_{S_2})=\frac{e^{Score_{S_1}}}{e^{Score_{S_1}}+e^{Score_{S_2}}}
\end{equation}
where $Score_{S_1},Score_{S_2}$ are the strength scores for $S_1,S_2$.\footnote{This model is equivalent to the Elo ratings system for chess players, under a simple transformation relating strengths and ratings \cite{aldous2017elo}.}  Given an observed sample of pairwise comparisons and their outcomes, inference using this model (typically via MLE) yields estimates of strengths. In our approach, we use the Bradley-Terry model to assign ``strength'' scores to sessions based on the pairwise comparison outcomes generated by the speech-based comparison prediction model: if the BPRS score for the individual in session $S_i$ is greater than the BPRS score for the individual in session $S_j$, then $S_i$ is the ``winner'' in that comparison.

Overall, then, our method has three components. First, supervised training of the model in Fig.\ref{fig:pairwise} to predict, not an individual-level severity score from a session, but a binary \emph{comparison} outcome when comparing two people's sessions. Second, comparison-outcome predictions for pairs of previously unseen test sessions. And third, use of the Bradley-Terry model to infer strength scores for the individuals in the test population. The outcome of the last step is a ranking of the individuals in the test population. Under the limited-resource assumptions that motivated this study, the approach can be considered successful to the extent that individuals with higher ground-truth BPRS scores are ranked higher.

\vspace{-0.3cm}
\section{Experiments}

Experiments were conducted using 3-fold cross-validation. Each fold contains data from subjects with schizophrenia and healthy controls with varying severity.

Hyperparameter tuning for the pairwise comparison prediction model was done through a grid search. The learning rate was selected from a set of \{5e-4,1e-4,5e-5\}, while the learning rate patience for the learning rate scheduler was selected from a set of \{25,50,75\}. When it comes to the hyperparameters of the model architecture, kernel size of the convolution layers were selected from a set of \{2,3,4,5\} and the pooling layer structure was also selected from a set of \{Average pooling, Max pooling\}. The best performing model architecture comprised of 2 sets of convolution layers with a kernel size of 5 and max pooling layers.

The best performing pairwise comparison prediction model was trained for 200 epochs with an AdamW optimizer with an initial learning rate of 1e-4. A ``Reduce-On-Plateau'' learning rate scheduler was used during the training process, which had a learning rate patience of 50 epochs with a 0.01 threshold set on the validation accuracy of the model. The models were trained with cross-entropy loss.

The predicted outcomes of all test-set pairwise comparisons were then used as input to Bradley-Terry model inference, yielding ``strength'' scores and therefore a ranking of all test items. Ranking performance was evaluated against ground-truth rankings based on individuals' BPRS scores using the Spearman rank correlation coefficient $(\rho)$ and Normalized Discounted Cumulative Gain (NDCG). 

Spearman correlations are widely used for ordinal data when the relationship is presumed monotonic but may or may not be linear.
NDCG, another widely used metric for evaluating ranking models, considers how well a predicted ranking aligns with a ground-truth ranking by considering both the relevance value of items and their positions in the list. NDCG is calculated by initially calculating Discounted Cumulative Gain (DCG). This sums relevance scores of the $n$ ranked items, applying a logarithmic discount to lower-ranked items to emphasize the importance of highly ranked results. 
    \begin{equation}
        DCG=\sum_{i=1}^{n}\frac{rel_{i}}{log_{2}(i+1)}
    \end{equation}
Here $rel_i$ is the ground truth relevance score (in our case the BPRS score) for the item at position $i$ in the model's predicted ranking.  Following that, Ideal Discounted Cumulative Gain (IDCG) is calculated. This represents the best possible ranking, where the most relevant items are ranked at the top. It is calculated using the same formula as DCG but with items sorted in descending order of relevance:
    \begin{equation}
        IDCG=\sum_{i=1}^{n}\frac{rel_{i}^{*}}{log_{2}(i+1)}
    \end{equation}
where $rel_{i}^{*}$ denotes the relevance scores sorted in the optimal order (ranked based on ground-truth relevance, here BPRS).
Finally NDCG is calculated by normalizing DCG by IDCG, resulting in a score between 0 and 1: 
    \begin{equation}
        NDCG=\frac{DCG}{IDCG}
    \end{equation}   

An NDCG score of 1 indicates a perfect ranking, while lower values indicate deviations from the optimal ranking. 

NDCG and Spearman rank correlation offer complementary insights, providing a more comprehensive evaluation when used together. NDCG prioritizes accurate placement of top-ranked items, making it ideal for scenarios where higher-ranked cases receive more attention. In contrast, Spearman correlation assesses overall rank consistency, measuring alignment with ground truth regardless of position. Together, they ensure the system both highlights the most critical items and preserves the overall ranking structure.

The rank ordered performance of the models is evaluated on the unseen test set. We report mean performance across all three folds. In addition to these experiments, we performed an ablation study to see how well our model performs on smaller subsets of the dataset, when compared to traditional regression-based severity estimation models. For that we used 90\%,80\%, and 70\% of the original dataset and trained and evaluated both our model and regression-based severity estimation model in \cite{gowtham_spade} with 3-fold cross validation on the smaller dataset.

\vspace{-0.3cm}
\section{Results \& Discussion}
\vspace{-0.3cm}

\begin{table}[h!]
  \caption{\centering
  \textbf{Rank-based performance comparison of the models}}
  \label{tab:rank-performance}
  \centering
  \begin{tabular}{ l c c c}
    \toprule 
    {\textbf{Model}}&Data Percentile&{\textbf{$\rho$}}&\textbf{NDCG}\\
    \midrule
    Regression model \cite{gowtham_spade}&100&\textbf{0.6876}&0.6073\\
    \textbf{Pairwise model (ours)}&100&0.6766&\textbf{0.6600}\\
    \midrule
    Regression model \cite{gowtham_spade}&90&0.6429&0.4947\\
    \textbf{Pairwise model (ours)}&90&\textbf{0.6465}&\textbf{0.6242}\\
    \midrule
    Regression model \cite{gowtham_spade}&80&0.6129&0.4713\\
    \textbf{Pairwise model (ours)}&80&\textbf{0.6553}&\textbf{0.6280}\\
    \midrule
    Regression model \cite{gowtham_spade}&70&0.6286&0.5279\\
    \textbf{Pairwise model (ours)}&70&\textbf{0.6378}&\textbf{0.5539}\\
    \bottomrule
    \end{tabular}
    \vspace{-0.6cm}
\end{table}

Table \ref{tab:rank-performance} summarizes the performance of our proposed approach using rank-based evaluation metrics, compared to a state-of-the-art regression model for symptom severity estimation. The results show that our pairwise comparison model consistently outperforms the regression model on NDCG across all dataset variants, demonstrating its superior ability to prioritize subjects with higher severity. In terms of Spearman's rank correlation, our model achieves better performance in three out of the four test cases, with only a slight decrease observed in the 100\% test set. These results indicate that the pairwise approach not only maintains strong overall rank consistency but also excels in high-severity case identification.

The consistently better performance across both metrics highlights the effectiveness of our method, even in low-resource scenarios simulated using dataset subsets. Moreover, the model continues to match or exceed the performance of the regression baseline on the full test set, reinforcing its robustness and suitability for real-world clinical prioritization.

An error analysis was conducted on sessions ranked in the upper tertile of the full test set. On average, the regression model misclassified 3.67 sessions as falling outside the upper tertile, whereas the pairwise comparison model misclassified only about one session. These results demonstrate that the pairwise comparison model is more effective than the regression model at prioritizing sessions with higher severity.

\vspace{-0.3cm}
\section{Conclusion and Future Work}
\vspace{-0.3cm}
This paper presents a speech-based pairwise comparison framework to support clinicians in prioritizing assessment and intervention for schizophrenia based on symptom severity. Leveraging articulatory and acoustic speech features, our approach applies the Bradley-Terry model to pairwise comparisons, yielding clinically meaningful severity rankings. Compared to conventional regression approaches, our method achieved consistently superior performance across multiple ranking metrics, underscoring its effectiveness in capturing relative symptom severity and prioritization of higher severity sessions. Notably, its stronger performance on smaller dataset subsets further highlights its robustness and potential utility in resource-constrained clinical settings where data may be limited.

Looking ahead, we aim to expand this work into a multimodal framework that incorporates speech, video, and text, building on recent advances in pairwise ranking using large language models. We also plan to explore ranking based on specific mental health subscales and evaluate the system on larger, more diverse datasets, including within prioritization-focused evaluation paradigms.

\bibliographystyle{IEEEbib}
\bibliography{refs}

\end{document}